\documentclass[nofootinbib,superscriptaddress]{revtex4}
\pdfoutput=1
\usepackage{amsmath,amssymb}
\usepackage{epsfig}
\usepackage{hyperref}
\usepackage{color}
\usepackage{slashed}
\usepackage{placeins}

\usepackage{amsfonts}
\usepackage{graphicx, rotating}
\usepackage{epstopdf}
\usepackage{latexsym}
\usepackage{rotating}
\usepackage[normalem]{ulem}

\usepackage[export]{adjustbox}

\usepackage[font={small}]{caption}   

\definecolor{commam}{rgb}{0.2,0.5,1.0}
\definecolor{myred}{rgb}{0.9, 0.2, 0.1}
\definecolor{myblue}{rgb}{0, 0.3, 0.7}
\definecolor{mygreen}{rgb}{0.04, 0.7, 0.5}

\hypersetup{colorlinks,citecolor=myred,linkcolor=myblue,urlcolor=myblue,linktocpage=true}

 \def\be   {\begin{equation}}   \def\ee   {\end{equation}}
 \def\ba   {\begin{array}}      \def\ea   {\end{array}}
 \def\bea  {\begin{eqnarray}}   \def\eea  {\end{eqnarray}}
 \def\bean {\begin{eqnarray*}}  \def\eean {\end{eqnarray*}}
 \def\nn{\nonumber}
 \def\bry{\begin{array}}
 \def\ery{\end{array}}

\numberwithin{equation}{section}

\begin{document}

\begin{flushright}
DESY 17-145
\end{flushright}

\vspace{1.5cm}

\title{
Light Higgs Boson from a Pole Attractor  
}

\author{Oleksii Matsedonskyi}
\email{oleksii.matsedonskyi@desy.de}
\address{DESY, Notkestra\ss e 85, D-22607 Hamburg, Germany}
\author{Marc Montull}
\email{marc.montull@desy.de}
\address{DESY, Notkestra\ss e 85, D-22607 Hamburg, Germany}

\begin{abstract}
We propose a new way of explaining the observed Higgs mass, within the cosmological relaxation framework. 
The key feature distinguishing it from other scanning scenarios is that the scanning field has a noncanonical kinetic term, whose role is to terminate the scan around the desired Higgs mass value. We propose a concrete realization of this idea with two new singlet fields, one that scans the Higgs mass, and another that limits the time window in which the scan is possible. Within the provided time period, the scanning field does not significantly evolve after the Higgs field gets close to the Standard Model value, due to particle production friction.
\end{abstract}

\maketitle


\tableofcontents

\newpage

\section{Introduction}

One of the main remaining puzzles of the Standard Model (SM), the Higgs mass, led physicists to search for heavy electroweak (EW) charged new physics at the TeV scale, as predicted by various scenarios, such as supersymmetry and composite Higgs. An alternative approach to the problem, named cosmological relaxation~\cite{Graham:2015cka} (see \cite{Espinosa:2015eda,Hardy:2015laa,Antipin:2015jia,Batell:2015fma,Matsedonskyi:2015xta,Evans:2016htp,Hook:2016mqo,Lalak:2016mbv,Tangarife:2017rgl,Batell:2017kho,You:2017kah,Nelson:2017cfv,Huang:2016dhp,Fowlie:2016jlx,DiChiara:2015euo,Marzola:2015dia} for subsequent developments), does not, {\it a priori} require this to be the case and can make the new physics either too heavy and beyond the reach of the current colliders, or very light and very weakly coupled. Given this difference, it seems especially important to examine theoretically this new concept to the greatest possible extent.  

The key ingredient of cosmological relaxation scenarios is the coupling of the Higgs to a new spin-zero field, the relaxion. This coupling induces the Higgs mass dependence on the relaxion field value. Cosmological evolution of the latter then leads to the Higgs mass scan, starting from some generic large value, down to the much smaller value which is currently observed. The scan is stopped in the right place due to a backreaction of the Higgs on the relaxion evolution.  
Existing realizations of this mechanism feature a scalar potential characterized by two hierarchically different periods. The larger period is needed for the complete Higgs mass scan and the smaller one allows to settle the final Higgs mass at the EW scale. The task of producing a UV completion for such a potential is very nontrivial and requires a dedicated model building~\cite{Choi:2015fiu,Kaplan:2015fuy,McAllister:2016vzi,Fonseca:2016eoo}. Interestingly, another known type of scanning scenarios, proposed in~\cite{Dvali:2003br,Dvali:2004tma}, does not require this feature. Instead of producing the short-period potential barriers for the scanning field, the whole scanning sector effectively decouples from the Higgs sector close to the SM Higgs mass. In the same spirit acts the mechanism proposed in this work.

We will examine a possibility of the Higgs mass scan termination by a noncanonical kinetic term of the relaxion field $\phi$. 
For this to happen we will assume that the field-dependent prefactor of $(\partial_\mu \phi)^2$ starts growing when the Higgs mass approaches its SM value. Enhancement of the kinetic term coefficient then results in the effective suppression of the relaxion potential and its coupling to the Higgs boson. With enough suppression the Higgs mass scan can slow down to an unobservably small speed. The relaxion field gets frozen around the value which gives the desired Higgs mass, which we will call an attractor point.         
Throughout this paper we will discuss concrete ways of implementing this idea. The main model-building challenge lies in finding a proper way to connect the value of the relaxion kinetic term with the Higgs mass. 
We start our analysis with a toy model, featuring a Higgs-dependent relaxion kinetic term $\sim 1/h^n (\partial_\mu \phi)^2$. Although this model straightforwardly realizes the $\phi$ kinetic term growth at small Higgs vacuum expectation values ({\it vev}s), it turns out to be incapable of producing a naturally light SM-like Higgs. But its detailed analysis proves to be useful in explaining some basic features of relaxation with noncanonical kinetic terms and, more importantly, provides a guideline for constructing realistic models. 
We present one such a model in the following, which features an extra scalar field $\chi$. It is now this new field which is responsible for the growth of the relaxion kinetic term $\sim 1/\chi^n (\partial_\mu \phi)^2$. 
In our construction, the $\chi$ field is not sensitive to the Higgs field, and simply rolls for a fixed amount of time, until it reaches the pole value, where the relaxion evolution is effectively terminated. 
A desired sensitivity of the relaxion to the Higgs mass is achieved by the $h$-dependent particle production friction of the relaxion. This friction is initially absent, allowing the relaxion to scan the Higgs mass. Once the Higgs mass approaches its SM value, the friction turns on.  After that, the particle production significantly slows down the $\phi$ evolution, until the time it becomes completely shut down by the $\chi$-dependent kinetic term when $\chi$ finally approaches the pole.

The structure of this paper is the following. In Sec.~\ref{sec:toy} we start by introducing a toy model. In Sec.~\ref{sec:twofieldattr} we describe a more complex setup, with two singlet fields; we discuss its one-loop structure and its evolution before and after the Higgs mass scan. The details of the scan are discussed in Sec.~\ref{sec:scan}, preceded by a brief review of the particle production friction. Finally we discuss our results in Sec.~\ref{sec:disc}.

\section{Toy Model}
\label{sec:toy}

\subsection{Main idea}
\label{sec:toyidea}

As usual in the scanning scenarios, we promote the Higgs mass to a field-dependent variable by coupling the Higgs to another field, a spin-zero SM singlet $\phi$. While we assume the Higgs potential to take a generic form, controlled by a cutoff $\Lambda$, 
the interactions of the $\phi$ field are kept under control by imposing the shift symmetry $\phi \to \phi+c$, which is weakly broken by a dimensionless parameter $\kappa$. The leading terms of the resulting scalar potential are
\be\label{eq:scanpot}
V = \kappa \Lambda \phi h^2-\kappa \Lambda^3 \phi - \Lambda^2 h^2 + {\lambda} h^4 \,,
\ee
which makes the Higgs mass parameter depend on the $\phi$ {\it vev}
\be\label{eq:mh}
m_h^2=-\Lambda^2+\kappa \Lambda \phi \, .
\ee
Here and in the following we use $h^2$ for $h^\dagger h$, and for conciseness omit most of the order-one factors, as well as the Higgs quartic coupling constant $\lambda$ and the quartic coupling term itself.  In Eq.~(\ref{eq:mh}) we fixed the $\phi$-independent part of the Higgs mass term to be negative, and the starting $\phi$ value is chosen to be less than $\Lambda/\kappa$, ($\phi=0$ for simplicity) such that the Higgs {\it vev} is initially of a cutoff size\footnote{We will assume some, at least small, separation between the initial Higgs mass and the cutoff to make our effective field theory (EFT) analysis meaningful.}. 
To make the Higgs {\it vev} decrease and approach the SM value, we fixed the sign of the leading term of the $\phi$ potential $\kappa \Lambda^3 \phi$ so that $\phi$ increases with time. 
Notice that we have not required any fine-tuning of the theory parameters. As for the initial conditions, if we assume a uniform distribution of the relaxion field values over different space points in the beginning of the process, only an order-one fraction of them will give the $\phi$ value below $\Lambda/\kappa$. But despite not being able to assure the needed initial value, we are satisfied with an order-one probability for it. As the inflation stretches away the field inhomogeneities, we can assume our initial condition $\phi=0$ to hold soon after the beginning of inflation everywhere in the given causally connected part of the Universe. All the discussion so far closely followed the original relaxion proposal~\cite{Graham:2015cka}, up to the sign of the initial Higgs mass.

\medskip
As was anticipated in the Introduction, the core of our scenario is the mechanism allowing us to stop the scanning when the Higgs mass approaches the SM value, using the diverging $\phi$ kinetic term. For this toy model we will simply take the kinetic Lagrangian
\be\label{eq:kinlag} 
{\cal L}_{\text{kin}} = \frac{\Lambda_{\text k}^{2n}}{h^{2n}} (\partial_\mu \phi)^2 +  (\partial_\mu h)^2
\ee
where $\Lambda_{\text k}$ is a mass dimension-one parameter and $n$ is some positive power. 
For now we will not give any comments on possible UV completions producing this type of kinetic terms, and we ignore such questions as the naturalness of the kinetic term choice and the scalar potential.    
The main purpose of this section is to introduce the reader to the dynamics of the relaxion with a noncanonical kinetic term.  

 As can be immediately read off from Eq.~(\ref{eq:kinlag}), the prefactor of the $\phi$ kinetic term starts exploding upon approaching to $h=0$. This means that every additional unit $\phi$ variation takes more and more time, and at some point the $\phi$ evolution effectively stops, with a Higgs {\it vev} and mass being close to zero. 
Interestingly, the attractor point thus generated, $\phi =\Lambda/\kappa$ and $h=0$, does not correspond to either the local or global minimum of the $\phi$ potential. Clearly, in order to reproduce the SM we need the attractor to be around the SM Higgs {\it vev} $h=v$ and not at zero. In this introductory section we will however limit ourselves with a less realistic but simpler case.   

The metric on the field space described by the kinetic terms~(\ref{eq:kinlag}) is not flat; therefore we are not able to canonically normalize both fields in all the time points simultaneously. To have a first glance on the details of their evolution we will integrate out the Higgs field using $h^2 \to(\Lambda^2-\kappa \Lambda \phi)$~\footnote{This substitution is not always valid, as the Higgs potential can also receive contributions from the $\phi$ kinetic term. The regime of its validity will be discussed later on.}. We thus arrive at the one-field Lagrangian
\be\label{eq:onefield}
{\cal L}_{\phi}={\Lambda_{\text k}^{2n} \over (-\kappa \Lambda \phi)^{n}} (\partial_\mu \phi)^2 +\kappa \Lambda^3 \phi
\ee  
where, for simplicity, we performed a shift $\kappa \Lambda \phi -\Lambda^2 \to \kappa \Lambda \phi$. In new notations the   attractor point simply corresponds to $\phi = 0$ and the initial $\phi$ value is negative. 
We can now switch to a new canonically normalized field $\hat \phi$ so that 
\be
\frac{\Lambda_{\text k}^{n}}{(-\kappa \Lambda \phi)^{n/2}} \partial \phi = \pm \partial \hat \phi
\ee
Depending on the power $n$, we choose the following redefinitions (omitting obvious constant factors)
{\renewcommand{\arraystretch}{1.6} \begin{center}
\begin{tabular}{c | c | c}
$n=1\,$  &  
$\;-\phi \sim \hat \phi^2\;$ & $(-\infty,0) \to (-\infty,0)$ \\
\hline
$n=2\,$  &  
$\;-\phi \sim \exp [- \hat \phi]\;$ & $\;(-\infty,0) \to (-\infty,\infty)\;$\\
\hline
$n>2\,$  &  
$\; -\phi \sim  \hat \phi^{\frac{1}{1-n/2}}\;$ & $(-\infty,0) \to (0,\infty)$\\
\end{tabular} \end{center}}
where in the last column we showed how the interval $\phi \in (-\infty,0)$ maps onto the canonical field $\hat\phi$. It shows that the $n=1$ case is special as the attractor point is mapped onto a finite value of $\hat \phi$. 
This potentially causes a problem once we try to move the attractor point away from $h=0$ to some finite nonzero value. The simplest way to do so is by changing the $\phi$ kinetic term to $(\partial_\mu \phi)^2/(h^2-\Delta^2)^n$. Now, for $n=1$ the $h^2=\Delta^2$ corresponds to a finite $\hat \phi$. Hence it can be reached and also overshot, making the $\phi$ kinetic term negative.    

\begin{figure}
\centering
\includegraphics[width=.90\textwidth]{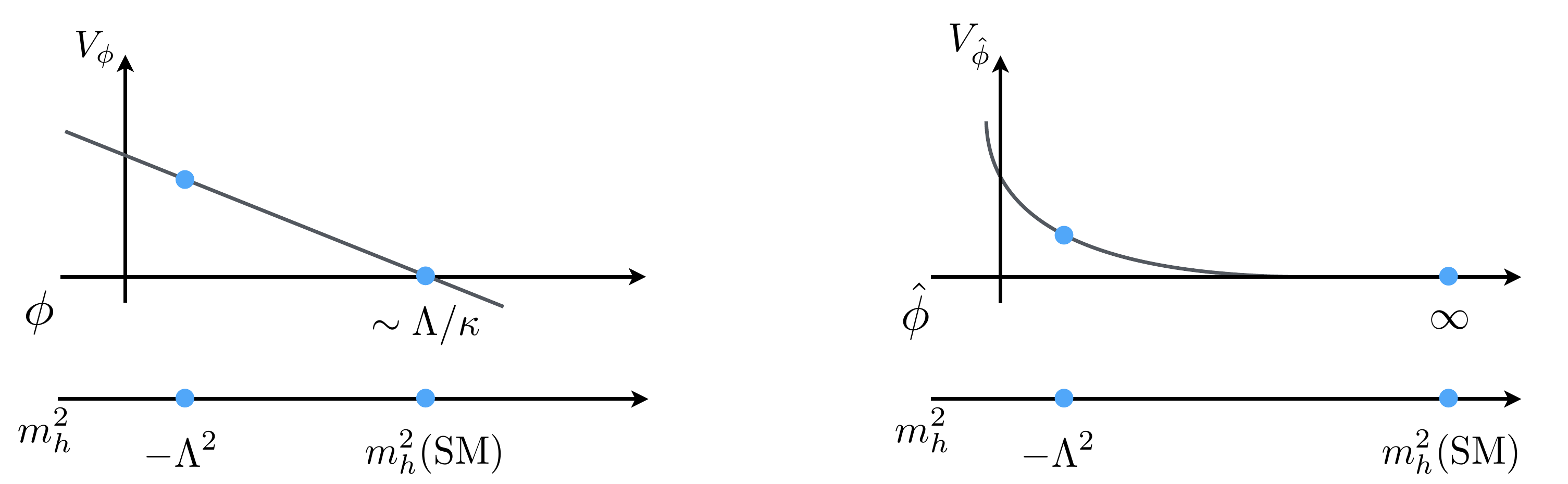}
\caption{Schematic representation of the scanning field potential and the corresponding Higgs mass, in terms of the initial field $\phi$ and after its canonical normalization, as explained in the text. 
}
\label{fig:1}
\end{figure}

For $n\geq2$, on the other hand, the attractor point is mapped to infinity and hence $\hat \phi$ will be eternally approaching it. This is because the stretching of the $\phi$ field outruns the $\phi$ time variation as we approach the attractor along the chosen trajectory.    
The resulting evolution of the $h-\phi$ system for $n\geq 2$ is schematically depicted on Fig.~\ref{fig:1}. 
It is interesting to notice that the same behavior which we observe for $\hat \phi$ after setting the Higgs into the minimum of its potential is used in various models of pole inflation~(see e.g. \cite{Broy:2015qna} and references therein), with a difference that the inflaton field evolution goes away from what we call the attractor point. 

We will now perform a slight refinement of the previous analysis, which will highlight a few additional features of the toy model. In particular we will see why the toy model cannot be used to produce a light SM-like Higgs boson.

\subsection{Nonperturbativity}
\label{sec:strong}

Let us jump straight to the major phenomenological problem of this model, namely the exploding $h - \phi$ coupling. In order to make it apparent we will split $\phi$, $\dot\phi$ and $h$ into classical and fluctuation parts by 
\be\label{eq:redefhphi}
\phi = \phi_0 + \delta \phi\,, \;\;\; \dot\phi = \dot\phi_0 + {\delta \dot\phi}\,, \;\;\; h = h_0+\delta h\,,
\ee
 where the zero subscript denotes a classical background at some time $t_0$. On top of this we will locally (at the time point $t_0$) canonically normalize the $\phi$ field fluctuations $\delta \phi \to \hat \phi \, h_0^{n}/\Lambda_{\text{k}}^n$. After these manipulations the kinetic terms of the field fluctuations are contained in 
\be\label{eq:kinlag1} 
{\cal L}_{\text{kin}} \supset 
\frac {h_0^{2n}}{h^{2n}} (\partial_\mu \hat \phi)^2 + (\partial_\mu h)^2
\ee  
while the scalar potential~(\ref{eq:scanpot}) becomes, omitting fluctuation-independent terms
\be\label{eq:scanpot2}
V = 
(\kappa \Lambda \phi_0- \Lambda^2) h^2 + \kappa \Lambda \frac{h_0^{n}}{\Lambda_{\text{k}}^n} \hat \phi h^2
-\kappa \Lambda^3\frac{h_0^{n}}{\Lambda_{\text{k}}^n} \hat \phi
\ee
We can now for instance estimate the amplitude of the Higgs decay into two $\hat\phi$'s. Close to the attractor ($h_0^2 \to 0$) the $\hat \phi$ potential is negligible and we can treat $\hat \phi$ as massless. The $h-\hat\phi$ interaction then arises from the first term of Eq.~(\ref{eq:kinlag1}).
The resulting non-SM Higgs decay amplitude 
\be
A_{h \to \hat \phi \hat\phi} \sim \frac {m_h^2}{h_0}
\ee
is expected to be sizable, far beyond the experimental bounds, and moreover ill behaving close to the attractor.

\subsection{Locking of the pole field}
\label{sec:lock}

\begin{figure}
\centering
\includegraphics[width=.60\textwidth]{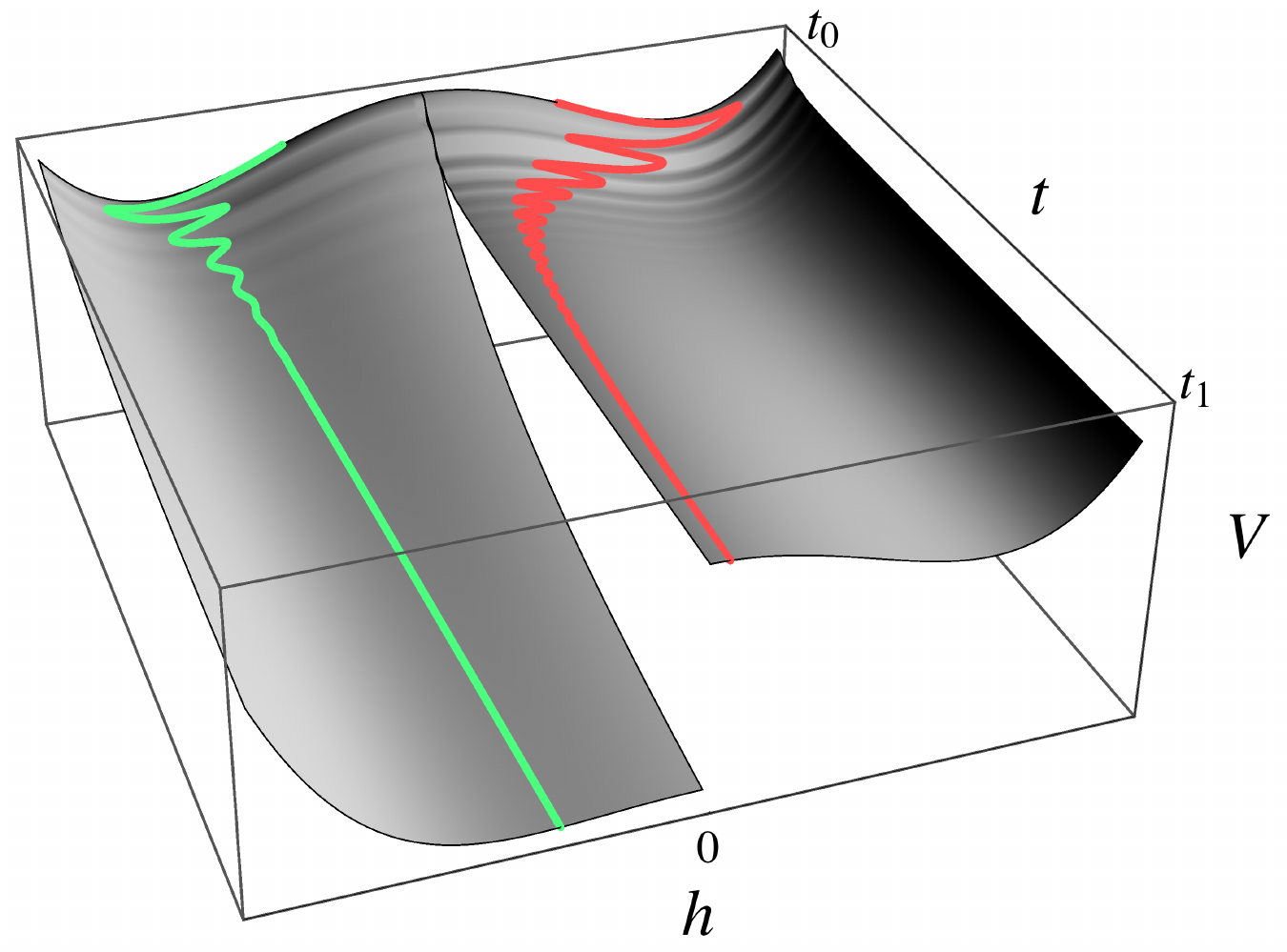}
\caption{Example of the Higgs field evolution without locking (left part, green line) and with locking (right part, red line). Values of the $\dot\phi_0$-independent part of the scalar potential $V$ are shown in grey. In the first case the Higgs field follows the $V$ minimum which evolves towards $h^2=0$, while in the second case the Higgs {\it vev} is driven towards zero, with the $V$ minimum frozen close to the initial value.}
\label{fig:2}
\end{figure}

Even though the previously discussed problem rules out the toy model, we will still make use of it to explain the locking of the $h$ field. It will be useful in the following as it also applies to any other field which produces the kinetic pole. 
After singling out the background component of the $\phi$ field~(\ref{eq:redefhphi}), the kinetic Lagrangian~(\ref{eq:kinlag}) also generates a $\dot \phi_0$-dependent term contributing to the Higgs potential
\be\label{eq:Vphidot}
\delta V = -\frac {\Lambda_{\text{k}}^{2n}}{h^{2n}}\dot \phi_0^2
\ee
which is minimized at $h^2=0$, thus competing with the rest of the Higgs potential which prefers $h^2 = \Lambda^2-\kappa \Lambda \phi$.
In order to understand when this extra term becomes important we need to find $\dot \phi_0$ by solving the equation of motion ({\it e.o.m.}) following from the Lagrangians~(\ref{eq:scanpot}) and~(\ref{eq:kinlag})
\be
\ddot{\phi_0} + \left( 3 \, H + n{ \, \partial_t (-h^2) \over  h^2}  \, \right) \dot{\phi}_0 = {h^{2n} \over \Lambda_{\text{k}}^{2n}} \,   V^\prime_\phi \,,
\ee
where $H$ is the Hubble parameter which appears after accounting for a metric expansion of the Universe. Besides the usual Hubble friction the equation contains a frictionlike term $\sim \partial_t h^2$ coming from the noncanonical form of the kinetic term. In order to estimate the maximal $\dot\phi_0$ we will consider the slow-roll limit, {\it i.e.} $\ddot \phi_0$ negligible compared to other terms. The maximal value of $\dot \phi_0$ is achieved when the friction is minimized, hence determined mostly by the irreducible Hubble expansion contribution  
\be
\dot{\phi}_{0\,\text{max}} \;\;\simeq\;\; {h^{2n} \over 3H \Lambda_{\text{k}}^{2n}}V^\prime_\phi
\ee 
Now substituting $\dot{\phi}_{0\,\text{max}}$ into Eq.~(\ref{eq:Vphidot}) we can estimate the maximal $\partial \delta V/\partial h$ and conclude that it is negligible compared to the potential~(\ref{eq:scanpot}) if 
\be\label{eq:locking}
\frac{(\kappa \Lambda)^2}{H^2} \, \frac{\Lambda^4}{\Lambda_{\text{k}}^4} \, \frac{h^2}{m_h^2} \, \frac{h^{2(n-2)}}{\Lambda_{\text{k}}^{2(n-2)}}<1 \,.
\ee 
In the opposite case $\delta V$ drives the Higgs $\it vev$ to zero, thus terminating $\phi$ evolution independently of the $\Lambda^2-\kappa \Lambda \phi$ value. 
We will call this termination process ``locking''. 
The evolution of the Higgs field with and without locking is shown in Fig.~(\ref{fig:2}). 
Therefore  $\delta V$ can significantly distort the evolution of the fields, and requires a special attention when considering this type of model. For the realistic model of the next section we will have to forbid this behavior for the relaxation to work.

\section{Formulation of a Two-Field Model}
\label{sec:twofieldattr}

The toy model analyzed in Sec.~\ref{sec:toy} was shown to fall into a strongly coupled regime close to the attractor, thus failing to reproduce the Standard Model.  
We will now show that a tractable realistic model of the pole attractor can be constructed using one additional spin-zero field $\chi$ which controls the relaxion kinetic term. 
The main goal of this section is to define the general structure of the two field ($\phi$ and $\chi$) model, while its detailed analysis and numerical results will be presented in Sec.~\ref{sec:scan}.

\subsection{Formulation}
\label{sec:twofieldintro}

The discussion of Section~\ref{sec:toy} suggests that the Higgs field cannot be simply put in the $\phi$ field kinetic term denominator. Hence we will introduce another spin-zero singlet field $\chi$ to produce a pole in the $\phi$ kinetic term. This will allow for more freedom in choosing the pole field properties, in particular, we would like to make the $\chi$ kinetic term have the same type of pole as the one of $\phi$. In the following  we will consider
\be\label{eq:twopoles}
{\cal L}_{\text{kin}} = \frac{\Lambda_{\text{k}}^2}{\chi^2} \left\{ (\partial_\mu \phi)^2 +(\partial_\mu \chi)^2 \right\}
\ee  
This solves the problem of strong coupling pointed out in Sec.~\ref{sec:strong}. We recall that it arises from the ill-behaved expansion of the Higgs field around the classical value $h_0\to 0$:
\be
\frac{(\partial_\mu \phi)^2}{h^{2n}} \to \sum_k \left(\frac{\delta h}{h_0}\right)^k \frac{(\partial_\mu \phi)^2}{h_0^{2n}}
\ee
Resulting interaction terms remain divergent even after $\phi$ is normalized canonically, as each $\phi$ removes only $n$ powers of $h_0$. Thus we get rid of this problem by switching to $\chi$, which gets normalized as well, absorbing the remaining poles.
The resulting theory is tractable within the usual perturbative approach. 
Notice that from this argument we are not strictly required to have the same order of poles for $\chi$ and $\phi$, but we will stick to this option for definiteness. Additionally, having  even-order poles we are safe from the problem of negative kinetic terms. 
The structures of the type~(\ref{eq:twopoles}) are also interesting as they are frequently used and motivated in supergravity models of inflation (see e.g.~\cite{Scalisi:2016hvu} for a review)~\footnote{Notice that most of the relaxion models only address the little hierarchy problem, i.e. their cutoff $\Lambda$ is significantly below $M_{\text{Pl}}$. Therefore the presence of the physics capable of explaining $\Lambda \ll M_{\text{Pl}}$ and for instance featuring supersymmetry (such as in~\cite{Batell:2015fma,Evans:2016htp}) or Higgs compositeness (see~\cite{Batell:2017kho}) is necessary at the scales above $\Lambda$.}. In particular, the kinetic Lagrangian (\ref{eq:twopoles}) can be associated to the K\"ahler potential $K \sim -\Lambda_{\text{k}}^2 \log[\Phi+\bar \Phi]$ of a chiral superfield $\Phi$. The scanning fields then are linked to the bosonic components of $\Phi$ as $\text{Re} \Phi \sim \chi$ and $\text{Im} \Phi \sim \phi$ . 

\begin{figure}
\centering
\includegraphics[width=.60\textwidth]{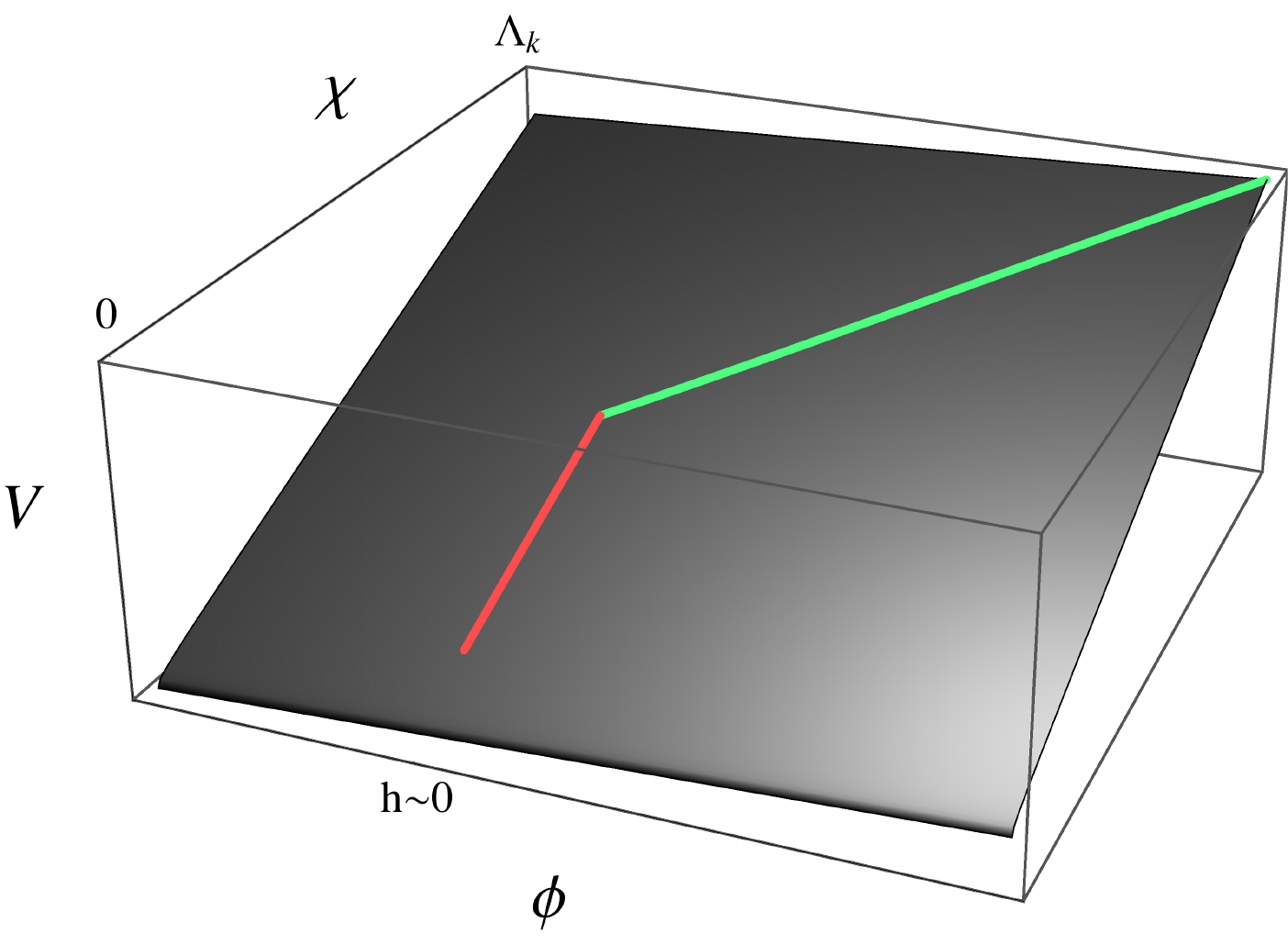}
\caption{Schematic plot of the $\phi-\chi$ system evolution. The evolution starts in the upper right corner. The change between the initial (green) and final (red) regimes happens around $h\sim 0$ due to the appearance of the particle production friction of the $\phi$ field. In the final point, both $\phi$ and $\chi$ evolution is effectively stopped by the growing kinetic terms.}
\label{fig:4}
\end{figure}

The next step would be to construct a mechanism which blocks the $\phi$ evolution as the Higgs mass becomes small. To achieve this, we will build a model that gives the following behavior. First, both $\chi$ and $\phi$ roll down their potential from the beginning of inflation, and the $\phi$ field scans the Higgs mass the same way as before. Second, as the Higgs {\it vev} reaches the SM value, the relative speed $\dot \phi/\dot \chi$ drops by a large factor. Therefore during the rest of the evolution until $\chi$ gets close to the pole, the field $\phi$ displaces by a much smaller amount than it did before the speed drop. As $\chi$ approaches the pole it blocks the $\phi$ evolution. The time dependence of the scanning fields is schematically shown in Fig.~\ref{fig:4}. In order to produce the relative speed drop we will use the Higgs-dependent particle production friction.
For instance, if $\phi$ has a particle friction, its time variation will be limited in the simplest case by $\dot \phi \sim H f$, where $H$ is the Hubble parameter and $f$ is a mass scale suppressing the particle production. In the absence of sizable particle production one will have instead a larger $\dot\phi$ controlled by the Hubble friction $\dot \phi \sim V^\prime_\phi / H$.     
We postpone the detailed discussion of different types of friction to Sec.~\ref{sec:partprod} and continue with the general description of the model.

Within the chosen approach, i.e. changing friction, we need to require that 
\begin{itemize}
\item the active scanning region (when $\chi$ is away from the pole) is long enough so the Higgs mass can be completely scanned;
\item after the Higgs mass reaches the SM value, $\dot \phi / \dot \chi$ has to decrease by at least a factor of $v^2/\Lambda^2$, so that $h$ mass is not changed significantly afterwards. 
\end{itemize}
In the following we will present a model where the $\dot \phi / \dot \chi$ drop originates solely from the growth of $\phi$ friction around $h=v$, while the $\chi$ friction is insensitive to $h$. The sole purpose of $\chi$ is thus to provide a limited time window for the scan, before it gets close to zero. This construction looks very different from the toy model. But the underlying principles, broadly defined, are similar -- the kinetic pole slows down $\phi$ when $h$ approaches $v$. The difference is that in the two-field model this backreaction on $\phi$ is delayed by the time $\chi$ needs to fall to the pole. This results in a certain amount of residual $\phi$ displacement which is suppressed by the high friction.

To fix the conventions we present the general (tree-level) Lagrangian of our model, omitting for the moment the terms relevant for the particle production, as well as the terms which are induced by quantum corrections
\bea\label{eq:twoflag}
{\cal L} &=& \frac{\Lambda^2_{\text{k}}}{\chi^2}\left\{ (\partial_\mu \phi)^2+(\partial_\mu \chi)^2 \right\} + \kappa_\phi \Lambda^3 \phi + \kappa_\chi \Lambda^3 \chi  \\
&& - \kappa_h \Lambda h^2 \phi  + \Lambda^2 h^2\,, \nn
\eea
where $\kappa$'s are positive dimensionless parameters. The choice of signs of different terms in the Lagrangian~(\ref{eq:twoflag}) already suggests that we will exploit the scanning with an initially large Higgs {\it vev} and a growing $\phi$ value. As usual, we will assume that the relaxion field changes by an amount $\sim \Lambda/\kappa_\phi$ during the scan, and fix for simplicity its initial value at $\phi=0$. We will also assume that $\chi$ starts negative and evolves towards the pole $\chi=0$.
Before analyzing the dynamics of the model we would like to address the stability of the relaxation mechanism against various possible modifications of Eq.~(\ref{eq:twoflag}), which could either be dictated by specific UV completions, or arise as quantum corrections in our EFT.  

First of all, the singular behaviour of the kinetic term guarantees that any additional contribution to the kinetic term with a weaker growth around the pole (i.e. weaker than $1/\chi^2$) can be neglected. Secondly, the mechanism is also insensitive to the displacement of the pole from the value $\chi=0$, which we choose for simplicity and also to make contact with the UV models discussed in \cite{Broy:2015qna}. For the mechanism to work, one only needs the $\chi$ field to roll towards the pole irrespectively of where exactly it is located. 
Given these considerations, after analyzing the quantum corrections to the effective action we do not find any contributions which can spoil the desired behavior of the kinetic terms. 

Let us now discuss the structure the effective potential. The tree-level scalar potential alone features shift symmetries for $\phi$ and $\chi$ in the limit of vanishing $\kappa_{\phi, h}$ and $\kappa_\chi$ respectively. 
Since the interactions described by the  $h^2 \phi$ term can induce the quantum corrections to the $h$-independent $\phi$ potential, we have to constrain $\kappa_\phi \gtrsim \kappa_h/(4 \pi)^2$. 
More importantly, unlike the kinetic Lagrangian, there are quantum corrections which can significantly alter the general form of the effective action and affect the field evolution for certain values of the fields and parameters. They come from the kinetic terms which explicitly break the $\chi$ shift symmetry. The $\chi$ potential therefore does not vanish even in the limit of $\kappa_\chi \to 0$. 
The analysis of these quantum corrections and their various implications is given in the two following subsections. Most importantly, we will show that these corrections are irrelevant for the scanning mechanism for a certain choice of model parameters.

\subsection{One-loop potential}
\label{sec:oneloop}

We will now discuss quantum corrections to the scalar potential, which are important for the general consistency of the mechanism and also specify the requirements to possible UV completions. 
This discussion will be qualitative and we will only obtain the general form of the most important quantum corrections omitting, in particular, terms dependent on $\dot \phi, \dot \chi \ll \Lambda^2$.
Before computing the loop corrections let us first switch to the field $\hat\chi$ defined as $\chi = - \Lambda_{\text{k}} \exp [-\hat\chi/\Lambda_{\text{k}}]$. The field range $\chi \in (-\Lambda_{\text{k}},0)$ is then mapped onto $\hat \chi \in (0,\infty)$. After this redefinition we obtain
\bea
{\cal L}_\chi &\;\;\;=\;\;\;  \frac{\Lambda_{\text{k}}^2}{\chi^2} (\partial_\mu \chi)^2 + \kappa_\chi \Lambda^3 \chi 
& \;\;\;\to\;\;\; (\partial_\mu \hat\chi)^2 - \kappa_\chi \Lambda^3 \Lambda_{\text{k}} e^{-\hat\chi/\Lambda_{\text{k}}} +\frac{\Lambda^4}{(4 \pi)^2} \hat\chi / \Lambda_{\text{k}} \label{eq:Lchi1}\\
{\cal L}_\phi &\;\;\;=\;\;\; \frac{\Lambda_{\text{k}}^2}{\chi^2} (\partial_\mu \phi)^2 +\kappa_\phi \Lambda^3 \phi
& \;\;\;\to\;\;\; e^{2 \hat\chi/\Lambda_{\text{k}}} (\partial_\mu \phi)^2 +\kappa_\phi \Lambda^3 \phi\label{eq:1111}
\eea
Notice the presence in Eq.~(\ref{eq:Lchi1}) of a linear contribution to the $\hat \chi$ potential. It appears due to the quartically divergent integral from the change of the path integral measure
\be
{\cal D \chi} \exp[iS] \to {\cal D \hat\chi} \exp[iS + \text{Tr}\,{\log[\cal \partial \chi/{\cal \partial \hat\chi}}]] \,.
\ee
The kinetic term of the new field $\hat \chi$ no longer contains interactions, and can induce no quantum corrections, unlike the $\chi$ kinetic term. 
However the presence of shift symmetry breaking interactions, associated with the kinetic term of the $\chi$ field, causes the presence of $\kappa_\chi$-independent term in the scalar potential of $\hat \chi$.
In order to recover this term, we could have alternatively used the noncanonical variable $\chi$ and computed the one-loop potential generated by the interactions contained in the $\chi$ kinetic term (we will do this in the following for the interactions contained in the $\phi$ kinetic term). Therefore the last term of Eq.~(\ref{eq:Lchi1}) can be effectively seen as a one-loop contribution. 
This contribution, as we will see later, is crucial for the $\chi$ dynamics. Before discussing it, let us also compute the one-loop effective potential~\footnote{Our estimates of the two-loop corrections show that they are irrelevant given that the cutoff of our model can reach 50~TeV at the very most.} arising from the interactions described by the new Lagrangians~(\ref{eq:Lchi1}),~(\ref{eq:1111}). 
The most important one-loop contribution arises from the $\hat \chi - \phi$ interactions encoded in the $\phi$ kinetic term
\be\label{eq:1loop}
V^{\text{(1-loop)}} \sim \frac{\Lambda^4}{(4\pi)^2} \log \left[{\Lambda^2}e^{2 \hat\chi/\Lambda_{\text{k}}} +\dots \right]\,,
\ee
which depends linearly on $\hat \chi$ for large $\hat \chi$, similarly to the last term of Eq.~(\ref{eq:Lchi1}). 
And, consequently, this correction also does not disappear in the limit of vanishing $\kappa_\chi$, $\kappa_\phi$. 

In order to estimate the relative importance of these loop terms, compared to the tree-level potential, we differentiate both with respect to $\hat \chi$
\be
\frac{\partial_{\hat \chi} V_{\text{tree}}}{\partial_{\hat \chi}V_{\text{loop}}} =
\frac{(4 \pi)^2 \kappa_\chi \Lambda_{\text{k}}}{\Lambda} e^{-\hat\chi/\Lambda_{\text{k}}}
\ee
and we find that the tree-level potential can only be dominant when $\hat \chi\lesssim\Lambda_{\text{k}}$ and $(4 \pi)^2 \kappa_\chi \Lambda_{\text{k}} > \Lambda$. We will assume that these requirements are satisfied during the active phase of the Higgs mass scan, as we would like to have a control over the $\chi$ potential during it.
The former constraint $\hat \chi\lesssim\Lambda_{\text{k}}$ is also needed to ensure that the $\phi$ evolution during the scan is unsuppressed by a large kinetic term and weakly dependent on $\chi$. We can add here another condition that $\Lambda_{\text{k}}$ has to satisfy, namely 
$\kappa_\chi \chi < \Lambda$ and consequently $\kappa_\chi \Lambda_{\text{k}} < \Lambda$. This is needed to provide a convergence of the $\kappa_\chi \chi/\Lambda$ expansion of our effective field theory.
These constraints lead to $\Lambda_{\text{k}} \sim \Lambda / \kappa_\chi$, which we will assume in the following.

The one-loop correction only becomes important for $\hat \chi > \Lambda_{\text{k}}$, {\it i.e.} when the $\phi$ kinetic term is already enhanced and the Higgs mass scan is mostly ended. In the following we will discuss the effects of this correction on the evolution outside the scanning window, namely on setting the initial and final conditions for the scan.

\subsection{Final vacuum after the scan}

Depending on its sign the correction~(\ref{eq:1loop}) would either block the $\chi$ movement to the pole and thus spoil the mechanism, or make it move towards the pole even faster. Assuming the latter to be the case, the $\chi$ potential becomes unbounded from below. Any phenomenologically viable UV completion of this type of model will therefore be required to contain a mechanism regularizing the scalar potential in the vicinity of $\chi=0$. This can be done for instance by adding to the potential an extra piece with a different functional form than~(\ref{eq:1loop}) to balance it and produce a finite minimum of the $\chi$ potential close to the pole. As another option we could shift the kinetic pole by a small constant
\be
\frac{\Lambda_{\text{k}}^2}{\chi^2 + \epsilon^2} \left\{(\partial_\mu \chi)^2 +(\partial_\mu \phi)^2  \right\}\,.
\ee 
This shift defines the maximal enhancement of the kinetic term, and thus the minimal slope of the $\phi$ potential and the time variation of the Higgs mass. It is thus limited by
\be
\delta h^2 \simeq \kappa_h \Lambda \,\delta \phi \simeq \kappa_h \kappa_\phi \Lambda^4 \left(\frac{\epsilon}{\Lambda_{\text{k}}}\right)^2 t^2 \ll v_{\text{SM}}^2 
\ee  
which for the current age of the Universe $t\sim10^{41}\,\text{GeV}^{-1}$ gives $\kappa (\epsilon/\Lambda_{\text{k}})<10^{-39}/(\Lambda/\text{GeV})^2$, for $\kappa=\kappa_\phi=\kappa_h$. Such a correction will not affect any details of the scanning mechanism and hence we will not discuss it any further. As a consequence of such a regularization, $\chi$ can actually reach the minimum of its potential and stop its evolution there. 

Notice, that such a regularization would also help to address the problem of the EFT validity of the model  
close to the pole. This problem is related to the fact that the quantum correction~(\ref{eq:1loop}) grows with $\hat \chi$, and can eventually lead to the EFT breakdown. We estimate this breakdown condition from $|V^{\text{(1-loop)}}| \sim \Lambda^4$, which gives the maximal allowed $\hat \chi$ value
\be\label{eq:eftbd}
{\hat \chi/ \Lambda_k} \sim (4 \pi)^2\,.
\ee
However, once the regularization mechanism starts acting, the growth of $|V^{\text{(1-loop)}}|$ stops and the EFT breakdown may not occur. The regularization, changing the $\chi$ behavior or the pole structure, however, should not happen before $\chi$ approaches the pole by an amount which is sufficient to block the residual Higgs mass variation to the acceptable amount. The minimal sufficient value of $\hat \chi$ can be estimated from 
\be
\delta h^2 \simeq \kappa_h \Lambda \,\delta \phi \simeq \kappa_h \kappa_\phi \Lambda^4 \exp [-2{\hat \chi}/{\Lambda_{\text{k}}}] t^2 \ll v_{\text{SM}}^2\,, 
\ee 
which gives 
\be\label{eq:scanblock}
{\hat \chi/ \Lambda_k} > \log \frac{10^{39}}{\kappa (\Lambda/GeV)^2}  \,.
\ee
For the values of $\Lambda$ and $\kappa$ obtained in the numerical scan below, 
we conclude that the $\hat\chi$ value of the EFT breakdown~(\ref{eq:eftbd}) is larger than the value of $\hat \chi$~(\ref{eq:scanblock}) at which the regularizing mechanism is allowed to start acting. 
And once the mechanism of the type described above turns on, the $\chi$ field settles in the minimum of the scalar potential. Hence the system does not arrive at a state violating the EFT validity.

\subsection{Initial conditions}
\label{sec:init}

The kinetic poles make the volume of the field space increasingly ``stretched'' upon approaching $\chi=0$. Thus, assuming uniformly distributed initial values of the renormalized fields over different patches of the Universe before inflation, we would find that most of the patches have $\chi$ which is very close to zero, almost completely blocking any possible Higgs mass scan. Having $\chi$ of the order of $\Lambda_k$, which is needed for a successful scan, would instead correspond to a very tuned, nontypical initial condition. 
We would like to emphasize that this problem only arises if $\chi$ field values are indeed distributed with a weight defined by the size of the kinetic terms. To verify this assumption we would need to know the exact UV completion. It can also be the case that the UV complete theory automatically sets initial values of $\chi$ sufficiently far from the pole. 
We will now show that we do not necessarily need to rely on this latter possibility and there can be ways to successfully complete the scan even with the uniformly distributed values of the renormalized fields.

One of the ways to solve this issue would be in adding the second kinetic pole at $\chi \lesssim - \Lambda_{\text{k}}$. It would stretch the field space at large $|\chi|$ and produce the second attractor value for the initial conditions. It will now be equally probable to start around the first or the second pole. Further, we require that the slope of the $\chi$ potential around this new pole repulse $\chi$ away from it towards zero. In this way, once $\chi$ starts its evolution close to the new pole, it will unavoidably pass the region $\chi \sim \Lambda_{\text{k}}$ where the scan can happen and then evolve to zero. If the scalar potential in the vicinity of poles is determined by its one-loop expression, its monotonic decrease with $\chi$ requires 
\be
V^{\text{(1-loop)}} \sim i \int \frac{d^4k}{(2 \pi)^4} \log \left[\frac{1}{(\chi-c_i)^2}+\frac{1}{\chi^2}\right] \; \rightarrow  \;
\begin{cases} 
\;-\Lambda^4 \log (\chi-c_i)^2 & {\chi \sim c_i} \\
\;\Lambda^4 \log \chi^2 & {\chi \sim 0}
\end{cases}
\ee
where $-c_i>0$ is the position of the extra pole. 
This sign-changing behavior may be achieved if the cutoff physics is sensitive to the {\it vev}s of the $\chi, \phi$ and $h$ fields, whose values change by an amount comparable to the cutoff during the evolution from one pole to another. 
An interesting consequence of such a construction is that $\chi$ becomes almost completely decoupled from all the other fields in the beginning and in the end of its evolution.
It is only active in the  window around $|\chi| \sim \Lambda_{\text{k}}$ when the Higgs mass scan happens. 

Alternatively, we could use the slowly varying $\chi$ field as the dominant source of inflation. In this case $\chi$ far from the minimum of its potential, and from the pole, would be a natural initial condition. In this case there is no need for the second pole, but a detailed analysis of such a possibility lies beyond the scope of this paper.

\section{Relaxation in the Two-Field Model}
\label{sec:scan}

\subsection{Review of particle production friction}
\label{sec:partprod}

As we have estimated in Sec.~\ref{sec:twofieldattr} for the two-field attractor, we need to produce an order $v^2/\Lambda^2$ drop of the ratio $\dot\phi/\dot\chi$ when the Higgs mass approaches the SM value. This section is dedicated to the brief review of the process allowing for this drop -- the particle production friction. The results given here are mostly based on works~\cite{Choi:2016kke,Hook:2016mqo,You:2017kah,Tangarife:2017rgl} where the particle friction was applied to the relaxion models, and the original model of inflation with a particle production~\cite{Anber:2009ua}. We would  like to emphasize that in the following we will be relying on analytic estimates of the relaxation dynamics. A comprehensive numerical study, while being important, lies beyond the scope of this paper. The results presented in this section will be applied to the two-field attractor dynamics in Sec.~\ref{sec:twofevol}.

We will consider an Abelian field $A_\mu$ with a mass $m_A$ coupled to one of the scanning fields ({\it e.g.} $\phi$ for definiteness) by means of an interaction
\be\label{eq:phiff}
{\cal L} \supset \frac{\phi}{f} F_{\mu \nu} \tilde F^{\mu \nu}
\ee
where $F_{\mu \nu}$ is the corresponding field strength tensor and $\tilde F^{\mu \nu}$ its dual. 
In the time-dependent $\phi$ background the transverse components of the $A$ field can acquire exponentially growing modes, draining the $\phi$ field kinetic energy, the process called ``particle friction''. To see how it appears we first write down the solutions of the {\it e.o.m.} for two transverse polarizations of $A$, derived using the WKB approximation~\cite{Choi:2016kke}
\be \label{eq:disp}
A[k]_\pm \sim \frac{1}{\sqrt{\omega_\pm}}e^{ - i \int d\tau \omega_\pm }\;\;\;\;\;{\text{with}}\;\;\;\;\; \omega_\pm^2 = k^2 \mp a k \dot \phi/f +  a^2 m_A^2 + a^2 \Pi_t[\omega_\pm,k]
\ee  
where $\pm$ stand for right and left helicity, $a$ is a scale factor of the expanding Universe, $\tau$ is a conformal time $a d \tau = d t$, and $k$ is a 3-momentum. The approximation~(\ref{eq:disp}) is valid for $|\partial_\tau\omega/\omega^2|<1$. Given that we are looking for exponentially growing gauge fields, one can end up in a space filled with a plasma of particles charged under $A$. Therefore in Eq.~(\ref{eq:disp}) we have also included the thermal correction to the dispersion relation $\Pi_t$~\cite{Bellac:2011kqa,Kapusta:2006pm}
\be
\Pi_t[\omega,k] = m_D^2 \frac{\omega}{k}\left(\frac{\omega}{k} + \frac 1 2 \left(1-\frac{\omega^2}{k^2} \right) \ln \frac{\omega + k}{\omega-k} \right) 
\ee 
with $m_D^2 = g_A^2 T_{\text{p}}^2/6$ defining the Debye mass of a plasma with a temperature $T_{\text{p}}$. 
If the dispersion relation~(\ref{eq:disp}) allows for imaginary $\omega \equiv i \Omega$, the vector field can experience exponential growth with time.  
Let us first notice that $\Pi_t$ is a positive function for complex $\omega$; therefore the existence of complex solutions of ~(\ref{eq:disp}) for one of the two polarisations requires
\be \label{eq:PF1}
a \dot \phi > 2 m_A f\,,
\ee
where without loss of generality we have assumed $\dot \phi>0$. 
We further notice that in the $\Omega \sim k$ limit $\Pi_t$ saturates around $m_{\text{D}}^2$ while for $\Omega \ll k$ we obtain $\Pi_t \sim m_{\text{D}}^2 |\Omega/k|$. Therefore, once the condition~(\ref{eq:PF1}) is satisfied, the maximal $\Omega$, and hence maximal instability, is given by 
\be \label{eq:PF2}
\Omega_{\text{max}} \sim 
\begin{cases}
a \dot \phi / f & \text{ for }\; a \dot \phi>2 m_D f \\
a \dot \phi / f \frac{(\dot \phi / f)^2}{m_{\text{D}}^2} & \text{ for }\; a \dot \phi \ll m_D f
\end{cases}
\ee
which shows that the instability growth becomes weaker in plasma. In all the cases the instability is maximized around 
\be
k_{\text{max}}\simeq a \dot \phi/2f
\ee 
We hence identified three regimes of $A_\mu$ evolution: with no exponential growth, with a fast growth, and with a growth slowed down by the high temperature plasma, $T_{\text{p}} \gtrsim \dot \phi/f$. In case of growing instability we can expect 
\be\label{eq:expprod}
\langle F_{\mu \nu} \tilde F_{\mu \nu} \rangle \sim \langle F_{\mu \nu} F_{\mu \nu} \rangle \sim \langle A_\mu^2 \rangle \sim H^4 \exp\left[2 \int d \tau \Omega_{\text{max}}\right]\,.
\ee
Notice that the time of the efficient exponential growth is limited by roughly one Hubble time, as after that the produced gauge field modes become significantly redshifted. In Appendix~\ref{app:pp} we collect the precise expressions for the quantities listed in Eq.~(\ref{eq:expprod}).
Thus produced energy density stored in the gauge field can lead to several effects.

\begin{itemize}

\item First, growing gauge field modes backreact on the rolling field $\phi$. The $\phi$ {\it e.o.m.} reads 
\be\label{eq:eomppf}
\ddot \phi+3H \dot \phi - \frac{1}{f} \langle F_{\mu\nu}\tilde F^{\mu\nu}\rangle + V^\prime_\phi=0\,.
\ee
For the sake of this section we forget about the noncanonical kinetic terms, as the scan happens around $|\chi| \sim \Lambda_{\text{k}}$, {\it i.e.} in the regime where the kinetic terms are not significantly enhanced. Hence our results will remain parametrically accurate. 
In case of negligible $\langle F_{\mu\nu}\tilde F^{\mu\nu}\rangle$ the $\phi$ evolution is driven by the slope of the potential $V$ and the Hubble friction, with a maximal speed defined by
\be
3H \dot \phi + V^\prime_\phi=0 \;\;\; \Rightarrow \;\;\; \dot \phi = -\frac{V^\prime_\phi}{3H}\,.
\ee
If instead the term $\langle F_{\mu\nu}\tilde F^{\mu\nu}\rangle$ dominates, one can reach a stationary regime with the evolution defined by the last two terms of Eq.~(\ref{eq:eomppf}):
\be\label{eq:partfrictbal}
 -\frac{1}{f a^4} \langle F_{\mu\nu}\tilde F^{\mu\nu}\rangle + V^\prime_\phi=0\,.
\ee
Using the exact dependence of $\langle F_{\mu\nu}\tilde F^{\mu\nu}\rangle$ on $\dot \phi$ we can estimate  the maximal $\dot \phi$ as
\be\label{eq:phidot}
\dot \phi \simeq f H \left(\frac{\text{max}[H, T_\text{p}]}{H} \right)^{2/3}
\ee
To find out which of the two types of friction dominates we simply need to check which of the two gives the minimal $\dot \phi$.

\item Second, if the gauge field $A_\mu$ couples to the Higgs boson at tree level, through $g_A^2 A^2 h^2$, it can lead to a restoration of the electroweak symmetry due to the effective temperature which we define as $T_{\text{eff}}^2 \equiv \langle A_\mu ^2 \rangle$. Therefore we do not need the Higgs and the gauge field to enter thermal equilibrium for this mass correction to appear, and $T_{\text{eff}}$ simply describes the classical value the $A$ field.

\item Finally, the energy density stored in the gauge bosons can thermalise, leading to a creation of a thermal plasma.  This plasma can then slow down the gauge field growth as described above. For thermalization to occur we need it to be faster than the Hubble expansion rate. We will consider two plasma production channels, a perturbative pair production and a nonperturbative Schwinger production.

Efficient perturbative pair production of charged fermions happens if the typical gauge quanta energy is higher than the fermion mass $m_f$, and if the pair production cross section, enhanced by the large multiplicity of gauge quanta $N_\gamma$, is higher than the Hubble expansion rate~\cite{Ferreira:2017lnd}
\be\label{eq:condpert}
\Omega > m_f \;\;\;\text{and} \;\;\; N_\gamma > \frac{128\pi}{N_f g_A^4}\,,
\ee
where $N_f$ is the number of produced fermionic degrees of freedom and $g_A$ is a gauge coupling.
The plasma temperature can be estimated as an order-one fraction of the overall energy released by the rolling $\phi$ field in one Hubble time, $T_{\text{p}}^4 \sim V^\prime_\phi \dot \phi H^{-1}$.

The nonperturbative Schwinger production is characterized by
\be
\Gamma= \frac{(g_A E)^2}{4 \pi^3H^3} \exp \left[-\frac{\pi m_f^2}{g_A E} \right]\,,
\ee
where $E$ stands for a modulus of the electric field analogue for $A$. 
For this production channel to be efficient one needs
\be\label{eq:condschw}
g_A E \gg m_f^2, H^2  \,.
\ee
And the maximal plasma temperature can be estimated as an order-one fraction of energy stored in the electric field $T^4_{\text{p}} \simeq E^2$~\cite{Tangarife:2017rgl}.

Once one of the fermion production channels opens, $A$ field modes can thermalize. Thermalization will happen if there is enough time for $A$ modes to interact with the plasma before they exit the horizon. Concretely, we require the mean free path of gauge quanta to be less than the Hubble length~\cite{Tangarife:2017rgl}
\be
g_A^4 T_{\text{p}}> (4 \pi)^2 H\,.
\ee

\end{itemize}

We now have at our disposal three different friction regimes --Hubble friction, thermal particle friction, nonthermal particle friction-- with a potentially significant relative speed change.  We have also identified the criteria leading to switching between different regimes, which one needs to satisfy when the Higgs mass gets close to its SM value.

\subsection{Details of evolution and numerical results}
\label{sec:twofevol}

We now present a concrete realization of the general scanning scheme presented in Sec.~\ref{sec:twofieldintro}. In this realization we keep the $\chi$ field friction constant, which makes this field a spectator, whose only purpose is to give a finite amount of time for the relaxion field evolution, until $\chi$ reaches the pole. As the $\phi$ field rolls down, it initially only has a Hubble friction. $\phi$ couples to the SM gauge bosons, but the Higgs mass squared starts large and negative, making the EW gauge bosons heavy and forbidding the particle friction for $\phi$, until the Higgs {\it vev} becomes small enough. Once this happens, $\phi$ gets slow, and during the rest of the time, given by $\chi$, does not evolve significantly. The friction is produced by the term~\cite{Hook:2016mqo}
\be\label{eq:phiZZ}
\frac{\phi}{f} (g_2^2 W_{\mu \nu} \tilde W^{\mu \nu} - g_1^2 B_{\mu \nu} \tilde B^{\mu \nu})
\ee
where $W$ and $B$ correspond to the SM weak and hypercharge gauge bosons and $g_{1,2}$ are their gauge couplings. The combination of the field strengths in Eq.~(\ref{eq:phiZZ}) is chosen so that it does not contain a photon, which is massless during the whole scan, hence not sensitive to the Higgs mass change.  

Notice that non-Abelian gauge bosons develop the so-called magnetic mass in plasma $m_{M}\sim g_2^2 T_{\text{p}}$~\cite{Linde:1980ts}, which blocks the instability development, hence weakening the friction. This can happen to the $W$ and even the $Z$ boson, as in the broken EW phase it contains $W_3$. The way out in this case would be a restoration of the EW symmetry immediately after the friction starts, making the Abelian hypercharge boson a mass eigenstate, and hence not mixing with the state possessing $m_{M}$.  As will be discussed later, the restoration of the EW symmetry is also necessary for another unrelated reason.

As for $\chi$, we can either leave it with the Hubble friction only, or assign it some other particle friction. In the latter case the easiest would be to couple it to a dark photon $\chi X_{\mu \nu} \tilde X^{\mu \nu}/f_X$. 

We will assume the inflation to happen in the background of the relaxation process. One important advantage of this is the absence of $\chi$ locking (see below), which requires a sizable Hubble scale. Another advantage is that inflation continues after the relaxation and washes away all its by-products, such as thermal plasma, which simplifies the phenomenology.  We are now ready to consider the Higgs mass scan in detail and list the conditions needed for the described mechanism to work.

\begin{enumerate}

\item Initially EW gauge boson masses (which we collectively denote $m_W$) have to be too heavy to be produced, so that $\phi$ quickly scans the Higgs mass. The particle friction turns on when $m_W$ approaches the SM value, i.e. when the following condition is satisfied~\footnote{If the friction switches on earlier, the $\dot \phi$ will slowly decrease, tracking $m_W$ evolution, until $m_W$ gets to $H$, where the equilibrium $\dot \phi \sim f H$ is reached. Generically, $\dot \phi$ will remain at $f H$ for roughly the same amount of time as it had in the $m_W$ tracking regime. During this time the Higgs mass will still be scanned. To make the residual Higgs mass change negligible, $H$ is forced to be very small and, consequently, $A^2$ to be too high, washing away the sensitivity to $h\sim10^2$~GeV. }:
\be\label{eq:cond21}
\frac{\dot \phi_{\text{in}}}{f} \simeq 2 m_W\,,
\ee
where the initial velocity $\dot \phi_{\text{in}}$, acquired before the particle production turns on, is defined by the Hubble friction
\be
\dot \phi_{\text{in}} = \frac{V^\prime_\phi}{3 H}\,.
\ee

\item As soon as the condition~(\ref{eq:cond21}) is satisfied, we need to restore the EW symmetry. 
This makes the Abelian gauge boson $B$ a mass eigenstate, allowing the associated instability to develop without creation of the magnetic mass~\cite{Hook:2016mqo}~\footnote{This is also necessary to produce an abrupt drop of $m_W$ to allow for a quick $\dot \phi$ decrease. 
Otherwise $\dot\phi$ will track the decreasing $m_W$ until it falls below H, resulting in $v\sim H$ and $\dot \phi \sim v f$. So, in order to produce a large $\dot \phi$ drop, the particle friction will have to start at $\dot \phi \gg v f$, which cannot work, see previous footnote. Since it is not the only reason to enforce the EW symmetry restoration, we will not discuss it in more detail.}. The symmetry restoration can be produced immediately by the classical value of the $A$ field, or a bit later by plasma if it forms (see the second point of Sec.~\ref{sec:partprod}).
We find that once the condition~(\ref{eq:cond21}) is fulfilled, the $A$ contribution dominates over the others; therefore we require
\be
\langle A^2 \rangle  \simeq \frac{\langle F \tilde F\rangle}{H^2} \simeq \frac{V^\prime_\phi f}{H^2} \gtrsim v^2 / g^2
\ee

\item The resulting $\dot \phi$ drop should be at least $\sim v^2/\Lambda^2$. In this case there will be no significant residual scan of the Higgs mass after it approaches the SM value. From Eq.~(\ref{eq:phidot}), this gives a constraint
\be
\frac{\dot \phi_{({\text{fin}})}}{\dot \phi_{({\text{in}})}} \lesssim \frac{v^2}{\Lambda^2}
\;\;\;\Rightarrow\;\;\;
\frac{f H \, (\text{max}[T_{\text{p}},H]/H)^{2/3}}{V^\prime_\phi/3H} \lesssim \frac{v^2}{\Lambda^2}
\ee
This condition is crucial to eventually ensure the naturalness of the weak scale with respect to the cutoff $\Lambda$, and it represents a rather tight upper bound on the inflationary Hubble scale (see the end of this section). Therefore we do not improve on this point with respect to other relaxion models operating during inflation, which typically feature analogous constraints.

\item The condition previously given in Eq.~(\ref{eq:phidot}) actually depends on whether the SM plasma is created or not. In the the latter case $\phi$ reaches the equilibrium speed $\dot \phi \simeq f H$, in the former the velocity will be higher than that because the plasma decreases the friction efficiency. The fermion plasma can be formed in two ways, either from perturbative or nonperturbative production. Let us now consider under what conditions these plasma formation channels can be active. In the case when both production channels are efficient, the one which gives the higher plasma temperature will dominate.  For simplicity we will assume that $W$ gauge bosons are never exponentially produced because of the magnetic mass. In this way we will obtain a conservative estimate of the allowed parameter space.

\begin{itemize}

\item To allow for a perturbative production of fermions at $v=0$ at a plasma temperature $T_{\text{p}}$ we need (see Eqs.~(\ref{eq:condpert}) and (\ref{eq:app2}))
\be\label{eq:cond12}
\exp \left[\frac{8}{\pi^2}\frac{H^4}{m_{\text{D}}^4} \xi^6 \right] \gg 4 \pi^2  \frac{H^{4/3}}{m_{\text{D}}^{4/3}} \frac{128 \pi}{N_f g^4}
\ee
where $N_f\sim {\cal O}(100)$ counts the number of SM fermionic degrees of freedom, $m_{\text{D}}^2 = g^2 T_{\text{p}}^2/6$ and $\xi \equiv \frac{\dot \phi}{f H}$ is defined from the balance $V_\phi^\prime \simeq F \tilde F/f $ as (for $F\tilde F$ see Eq.~(\ref{eq:app2}))
\be\label{eq:cond13}
\exp \left[\frac{8}{\pi^2}\frac{H^4}{m_{\text{D}}^4} \xi^6 \right]  \simeq \frac{V^\prime_\phi f}{\frac{1}{2\pi^3} (H/m_{\text{D}})^2 H^4 \xi^7} \simeq  \frac{{2\pi^3} V^\prime_\phi f}{H^{4/3} m_{\text{D}}^{8/3}}
\ee
Thus, comparing the expressions~(\ref{eq:cond12}) and~(\ref{eq:cond13}) we conclude that the plasma temperature cannot grow above 
\be
T^{4/3}_{\text{max}} \simeq \frac{V^\prime_\phi f}{H^{8/3}}\frac{6^{2/3}N_f g^{8/3}}{256}
\ee
as the plasma will stop being produced otherwise. 
At the same time plasma cannot go above the equilibrium temperature, defined such that the order-one fraction of all the gauge field energy gets thermalized. The equilibrium plasma temperature can be estimated as the total energy density lost by the rolling $\phi$ field in one Hubble time $T_{\text{p}}^4 \sim \delta \rho \sim V^\prime_\phi \dot \phi/H$
\be
T_{\text{eq}}^4 \simeq \frac{m_{\text{D}}^{2/3}}{H^{2/3}} (V^\prime_\phi f)\,.
\ee

\item For the Schwinger production, the fermion plasma is produced if $g E>m_f^2$ ($m_f=0$ in the unbroken phase) and (see Eq.~(\ref{eq:app3}) for $E$)
\be
(g E)^2 \gg 4 \pi^3 H^4 \;\;\Rightarrow\;\; g^2 \frac{H^{2/3}}{m_{\text{D}}^{2/3}} (V^\prime_\phi f) \gg 4 \pi^3 H^4
\ee
which defines the maximal plasma temperature above which the production stops
\be
T^{2/3}_{\text{max}} \simeq \frac{6^{1/3} g^{4/3} V^\prime_\phi f}{4 \pi^3 H^{10/3}}
\ee
At the same time the plasma temperature is also limited by the energy stored in the electric field. We can find this equilibrium temperature from $T^2_{\text{p}} \simeq E$:
\be
T_{\text{eq}}^4 \simeq \frac{H^{2/3}}{m_{\text{D}}^{2/3}} (V^\prime_\phi f)
\ee

\item Once the fermion plasma is formed (either from perturbative or nonperturbative production), it can only backreact on gauge bosons and the Higgs field if $g^4 T_{\text{p}}\gtrsim (4 \pi)^2 H$.

\item The stability of plasma also requires the EW symmetry to remain unbroken when it forms, otherwise we obtain the magnetic mass blocking the instability and the plasma production. Hence either the plasma temperature should be high enough to restore the symmetry ($\text{min}[T_{\text{max}},T_{\text{eq}}]>v$) or it has to be restored by the $A^2$ (Eq.~(\ref{eq:app5})) correction
\be
A^2 > v^2/g^2 \;\;\Rightarrow\;\; \frac{H^{2/3}}{m_{\text{D}}^{2/3}}\frac{V^\prime_\phi f}{H^2} > v^2/g^2
\ee
It is easy to check that this is always the case if the other mentioned constraints are satisfied.

\end{itemize}

\item The scanning time window, given by $\chi$, should exist long enough, allowing $\phi$ to scan the entire Higgs mass range. The typical time $\phi$ needs for it is 
\be\label{eq:tphi}
t_\phi \sim \frac{\Lambda/\kappa_h}{\dot \phi} \sim \frac{H^2} {\kappa_\phi \kappa_h \Lambda^2} \frac{1}{H}\,,
\ee
while the time it takes $\chi$ to pass its natural field range is
\be
t_\chi \simeq \frac{\delta \chi}{\dot \chi} \simeq \left(\frac{H}{\kappa_\chi \Lambda} \right)^2 \frac{1}{H} \;\;\text{or}\;\; \left(\frac{\Lambda}{\kappa_\chi f_X} \right) \frac{1}{H}\,,
\ee
depending on whether it has a Hubble or a dark gauge boson friction respectively. 
Requiring $t_\phi \sim t_\chi$ we arrive at
\be
\kappa_\chi \sim (\kappa_\phi \kappa_h)^{1/2} \;\;\text{or}\;\; \kappa_\chi \sim \kappa_\phi \kappa_h \frac{\Lambda^3}{f H^2} 
\ee
depending on which type of friction we choose for $\chi$.
From the expression~(\ref{eq:tphi}) we can also derive the constraint on the duration of inflation, as it should last long enough to allow for a complete scan. The minimal required number of $e$-folds is given by the prefactor of $1/H$.

\item The velocity-dependent contributions to the $\chi$ potential (analogous to those discussed in Sec.~\ref{sec:lock}) will tend to push it to the poles, ruining its monotonic decrease (if we assume two poles as in Sec.~\ref{sec:init}), and significantly decreasing the scan window (if $\chi$ manages to start moving towards zero).   
To prevent $\chi$ locking we need the term $\frac{\Lambda_{\text{k}}^2}{\chi^2} \dot \phi^2$ to be subdominant for $\chi$ evolution compared to $\kappa_\chi \Lambda^3 \chi$. 

The maximal size of the former is reached for $\dot \phi \simeq \kappa_\phi \Lambda^3 / 3H$. Therefore we need
\be
\frac{\partial_\chi(\frac{\Lambda_{\text{k}}^2}{\chi^2} \dot \phi^2)}{\partial_\chi(\kappa_\chi \Lambda^3 \chi)} = 
\frac{\kappa_\phi^2 \Lambda^2}{9 H^2} \ll1
\ee 
where we took $\chi \sim \Lambda_{\text{k}} \sim \Lambda/\kappa_\chi$.

\item In order to make our classical consideration of the dynamics trustable, the Hubble expansion-induced field jumps have to be either absent or negligible compared to the classical field evolution. This gives the conditions 
\bea
v>H\,, &&\\
\delta \phi_{1H} \sim \dot \phi/H > H & \Rightarrow& f \text{max}[T_{\text{p}},H]^{2/3}> H^{5/3}\,, \\
\delta \chi_{1H} \sim \dot \chi/H > H & \Rightarrow& \kappa_\chi \Lambda^3 > H^3 \;\;\text{or}\;\; f_X > H\,,
\eea
where the $1H$ subscripts stand for a classical displacement in one Hubble time. 

\item Finally we have the constraint on the vacuum energy density, ensuring that inflation is not affected by our mechanism
\be
H^2 M_{\text{Pl}}^2 > \Lambda^4
\ee

\end{enumerate}

\begin{figure}
\centering
\includegraphics[width=.32\textwidth]{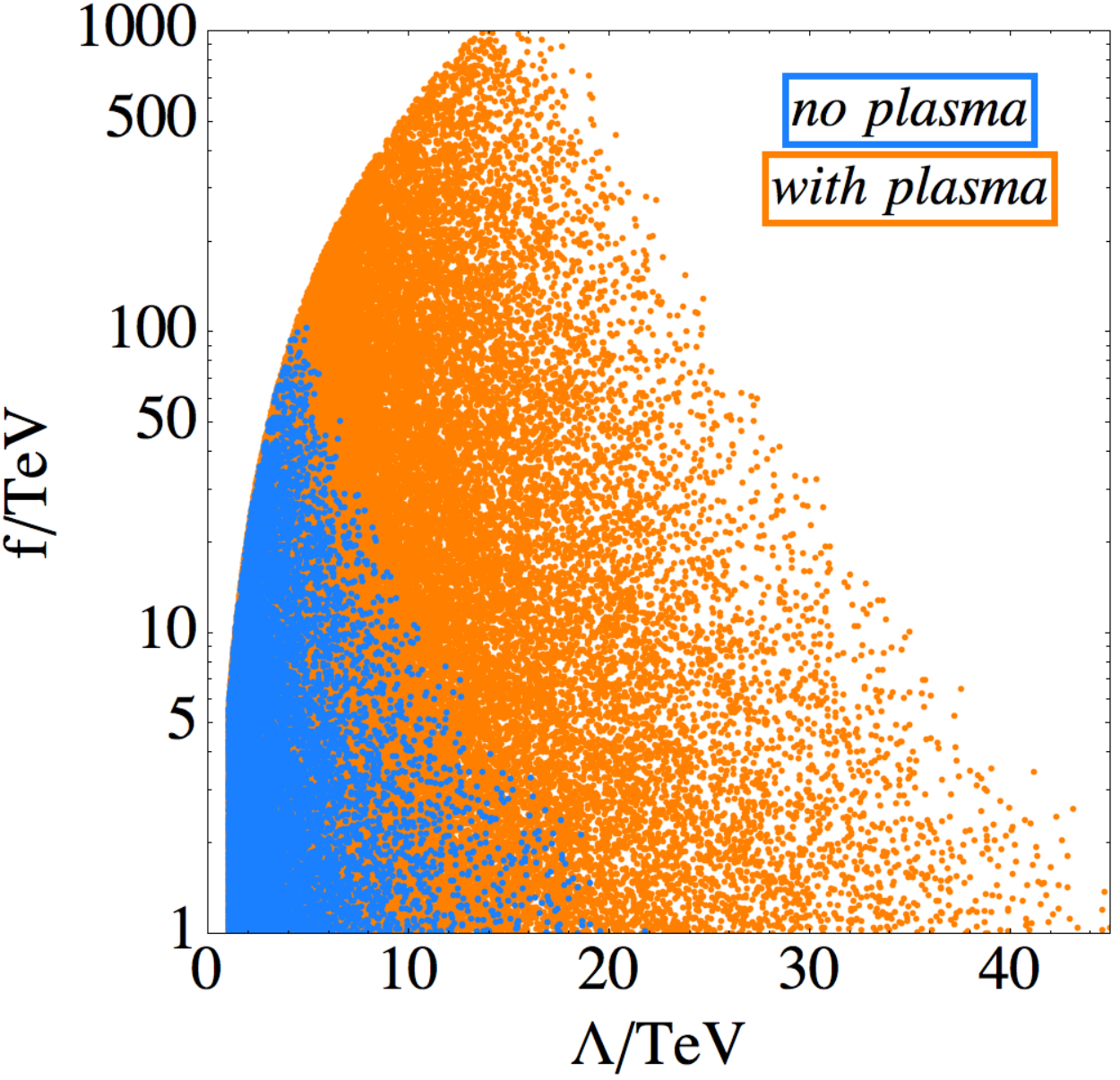}
\includegraphics[width=.32\textwidth]{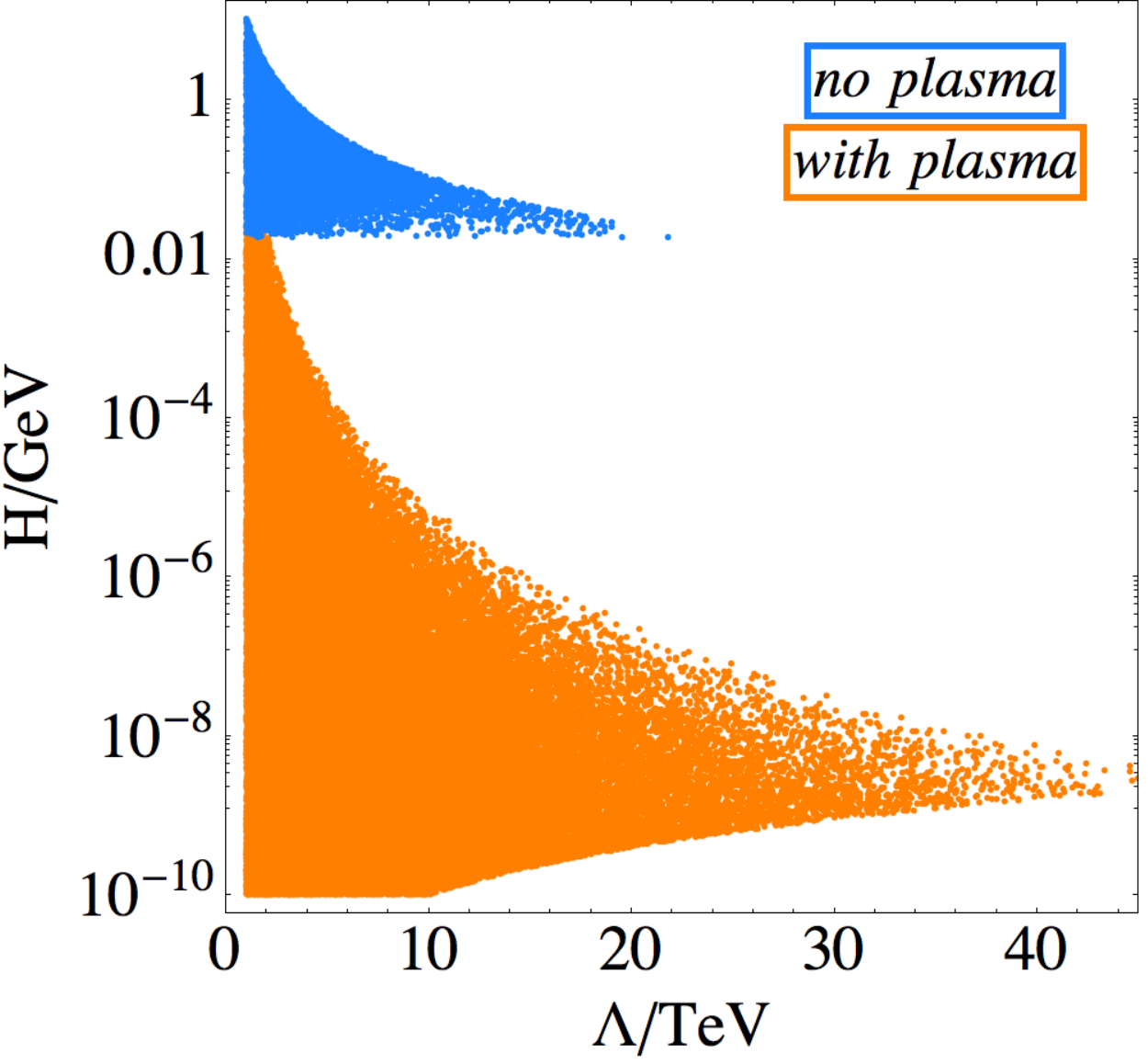}
\includegraphics[width=.32\textwidth]{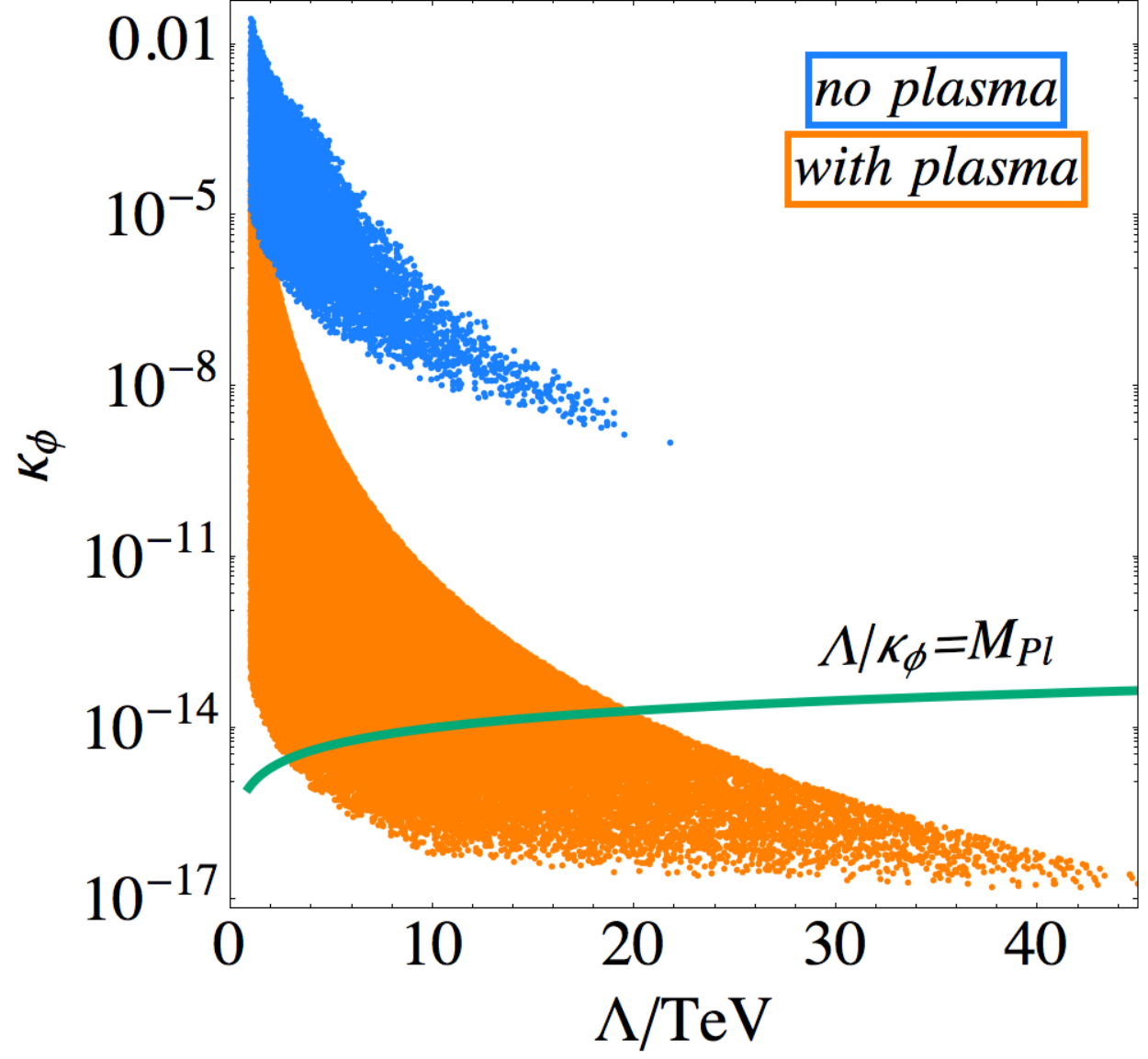}
\caption{Results of the numerical parameter space scan, in terms of $\Lambda$, $f$, $H$ and $\kappa_\phi$. Blue points correspond to evolution without plasma production, and orange to evolution with plasma production. The green line on the right plot shows minimal $\kappa_\phi$ below which the $\phi$ field excursion becomes trans-Planckian.}
\label{fig:3}
\end{figure}

The main result of this section is the scatter plots on Fig.~\ref{fig:3}, showing the values of the main relevant parameters which satisfy the constraints listed above. For all the points with the plasma production the main production channel is the perturbative production. The maximal allowed cutoff scale $\Lambda$ is around $50$~TeV, while without trans-Planckian field excursions it decreases to $\sim 20$~TeV. The friction without plasma is generated for a moderate (for this kind of scenario) Hubble scale ${\cal O}(0.1)$~GeV, while in the presence of plasma $H$ needs to be several orders of magnitude lower, with the maximal cutoff reached for $H\sim 10^{-8}$~GeV.

\section{Summary and Future Directions} 
\label{sec:disc}

We have discussed a new scenario with the dynamical Higgs mass scan, within a class of models pioneered in~\cite{Dvali:2003br,Dvali:2004tma} and~\cite{Graham:2015cka}. The novelty of our proposal lies in employing a noncanonical kinetic term of the scanning field, leading to its decoupling from the Higgs sector at the end of the scan. This scenario therefore does not require Higgs-dependent barriers for the relaxion (and hence new light EW-charged states), or, {\it a priori}, the presence of two periods in the relaxion potential.

We have presented a particular realization of the pole attractor idea, in which the scanning field can evolve only during a limited time window during inflation. After that the second field, controlling the kinetic terms, reaches the pole value and blocks the scan. Starting with a large Higgs {\it vev}, the scanning field first evolves quickly until the Higgs mass gets close to the SM value. At this point the scanning field evolution is slowed down by a dissipation of energy into SM gauge bosons, hence it displaces by a very small distance before the scanning window closes. We have identified two viable regions of the parameter space, with and without a backreacting plasma. The latter allows for a higher inflationary Hubble parameter, but a slightly lower maximal cutoff. 
The maximal cutoff of the Higgs sector $\sim 50$~TeV is well above the reach of current and near future collider experiments. It is nevertheless not restricted to be that far and can reside as low as the current lower bounds on new physics allow. The relaxion sector fields' couplings to the visible sector are exponentially suppressed making it very difficult to observe their direct effect. 
It is important to mention that this particular two-field realization of our scenario does feature two different ``periods" for the $\phi$ field, similarly to the original relaxion models. The first period is given by $\sim \Lambda/ \kappa_\phi$. The second periodicity $\phi\to \phi + 2 \pi f$ has to be assumed in order to ensure that the underlying physics responsible for the friction term~(\ref{eq:phiZZ}) does not induce additional unsuppressed shift symmetry breaking terms to the scalar potential.

All the models discussed in this work need the initial Higgs {\it vev} to be large, or in other words the Higgs mass squared to be negative. One can however think of modifications suitable to accommodate a vanishing initial Higgs field value $h=0$ and a large positive mass squared. These should include a change of signs in the scalar potential to provide a scan in the right direction. For the toy model, in order to prevent having a singularity in the relaxion kinetic term one would need to shift the pole away from $h=0$ as discussed in Sec.~\ref{sec:toyidea}, which in any case is needed to stop the scan at $h=v\ne0$. For the two field attractor one can think of assigning the Higgs-dependent particle production friction to $\chi$ instead of $\phi$, so that initially $\chi$ is slowed down by SM gauge boson production at $v=0$, and then the particle friction disappears and $\chi$ falls to zero, blocking the scan. The simplest implementation of this last mechanism however leads to a relatively low cutoff, at the TeV scale, so further refinements would be needed in order to increase it.

We would also like to mention that there is another potential way of implementing the two-field scenario, without using the concept of a limited scanning window and changing friction. This would be more in accord with the spirit of the toy model, where the pole depends on the Higgs {\it vev}. One can assign to $\chi$ a potential with a minimum (local or global) away from zero, which disappears or displaces to zero when the Higgs mass reaches the SM value. The simplest realization would be a scalar potential with a tadpole $V_\chi=h^2 \chi+\Lambda^2 \chi^2$, which however has the problem that the quantum correction $\sim \Lambda^4 \log \chi$ quickly becomes more important than the Higgs-dependent tadpole. It would be interesting to further investigate this model building direction. 

The proposed mechanism has a lot in common with the attractor models of inflation, sharing some structural features and, possibly, can find UV completions in analogue type of theories. In general,  the existence of an appropriate UV completion for our scenario presents an important question for further studies, in particular because the behavior close to the attractor relies on certain crucial  assumptions about the UV physics features.

Finally, we would like to mention that a distinctive phenomenological feature of this type of scenario is a slow change of the theory parameters with time, as the scan never completely stops. While we do not necessarily need this time variation to be detectable with the current experiments, its observation would be an interesting hint for this mechanism, especially given the limited direct experimental access to the relaxion sector, whose couplings to the SM particles are exponentially suppressed.

\vspace{5mm}
{\bf Acknowledgments}

We thank J. Elias-Mir\'o, C.Grojean, T. Konstandin, G.Servant and J.Wells for useful discussions and especially to A.Westphal and M.Scalisi for many important clarifications and suggestions. We would like to express a special thanks to the Mainz Institute for Theoretical Physics (MITP) for its hospitality and support during the completion of this work.

\newpage

\section{Appendix: Particle Production}
\label{app:pp}

The combinations of the field strengths can be rewritten in terms of electric and magnetic fields as~\cite{Anber:2009ua}
\be
\langle FF \rangle  \sim \langle E^2 \rangle + \langle B^2\rangle\;\;, \;\;\langle F\tilde F \rangle \sim \langle EB \rangle\,.
\ee
At zero temperature, they take the form~\cite{Anber:2009ua}
\bea
\langle E B \rangle & \simeq 10^{-4} H^4 \xi^{-3} \exp[ 2 \pi \xi]\,, &\label{eq:app2}\\
\langle E^2 \rangle &\simeq 10^{-4} H^4 \xi^{-3} \exp[ 2 \pi \xi] &\simeq \langle E B \rangle\,,\label{eq:app3}\\
\langle B^2 \rangle &\simeq 10^{-4} H^4 \xi^{-5} \exp[ 2 \pi \xi] &\simeq \xi^{-2}\langle E B \label{eq:app4} \rangle \,,
\eea
where we have defined $\xi \equiv \frac{\dot \phi}{f H}$.
Using the results of~\cite{Anber:2009ua} we can also estimate 
\be
\langle A^2 \rangle \sim \langle E B \rangle /H^2 \,.\label{eq:app5}
\ee
And the number of gauge field quanta in one Hubble volume as the total energy density $\sim k^2 A^2$ over the energy per particle $\Omega$~\cite{Ferreira:2017lnd}
\be
N_\gamma \simeq 10^{-4} \exp[4.5 \xi]\,.
\ee

In plasma we have~\cite{Tangarife:2017rgl}
\bea
\langle E B \rangle & \simeq \frac{1}{2\pi^3} \frac{H^2}{m_{\text{D}}^2} H^4 \xi^7 \exp\left[\frac{8}{\pi^2} \frac{H^4}{m_{\text{D}}^4} \xi^6\right]\,, & \\
\langle E^2 \rangle &\simeq \frac{1}{\pi^4} \frac{H^4}{m_{\text{D}}^4} H^4 \xi^9 \exp\left[\frac{8}{\pi^2} \frac{H^4}{m_{\text{D}}^4} \xi^6\right] 
& \simeq \frac{H^2}{m_{\text{D}}^2} \xi^{2} \langle E B \rangle 
\simeq \frac{H^{2/3}}{m_{\text{D}}^{2/3}} \langle E B \rangle \,, \\
\langle B^2 \rangle &\simeq \frac{1}{4\pi^2} H^4 \xi^5 \exp\left[\frac{8}{\pi^2} \frac{H^4}{m_{\text{D}}^4} \xi^6\right] 
& \simeq \frac{m_{\text{D}}^2}{H^2} \xi^{-2} \langle E B \rangle
\simeq \frac{m_{\text{D}}^{2/3}}{H^{2/3}} \langle E B \rangle\,,
\eea
where in the last column we used $\xi \sim (m_{\text{D}}/H)^{2/3}$, which is needed for the gauge field backreaction on $\phi$ to be sizable. We can also derive from the results of~\cite{Tangarife:2017rgl}
\be
\langle A^2 \rangle \sim \frac{1}{4 \pi^2} \xi^3 H^2  \exp\left[\frac{8}{\pi^2} \frac{H^4}{m_{\text{D}}^4} \xi^6\right] \sim \frac{H^{2/3}}{m_{\text{D}}^{2/3}} \langle E B \rangle/H^2\,,
\ee
and estimate the particle number
\be
N_\gamma \sim \frac{1}{4 \pi^2} \xi^2 \exp\left[\frac{8}{\pi^2} \frac{H^4}{m_{\text{D}}^4} \xi^6\right]  \sim \frac{1}{4 \pi^2} \frac{\langle E B \rangle}{H^4} \frac{H^{4/3}}{m_{\text{D}}^{4/3}}\,.
\ee



\begin{thebibliography}{99}

\bibitem{Graham:2015cka} 
  P.~W.~Graham, D.~E.~Kaplan and S.~Rajendran,
  Phys.\ Rev.\ Lett.\  {\bf 115}, no. 22, 221801 (2015)
  doi:10.1103/PhysRevLett.115.221801
  [arXiv:1504.07551 [hep-ph]].
  
\bibitem{Espinosa:2015eda} 
  J.~R.~Espinosa, C.~Grojean, G.~Panico, A.~Pomarol, O.~Pujol\'as and G.~Servant,
  Phys.\ Rev.\ Lett.\  {\bf 115}, no. 25, 251803 (2015)
  doi:10.1103/PhysRevLett.115.251803
  [arXiv:1506.09217 [hep-ph]].
  
\bibitem{Hardy:2015laa} 
  E.~Hardy,
  JHEP {\bf 1511}, 077 (2015)
  doi:10.1007/JHEP11(2015)077
  [arXiv:1507.07525 [hep-ph]].
  
\bibitem{Antipin:2015jia} 
  O.~Antipin and M.~Redi,
  JHEP {\bf 1512}, 031 (2015)
  doi:10.1007/JHEP12(2015)031
  [arXiv:1508.01112 [hep-ph]].
  
\bibitem{Batell:2015fma} 
  B.~Batell, G.~F.~Giudice and M.~McCullough,
  JHEP {\bf 1512}, 162 (2015)
  doi:10.1007/JHEP12(2015)162
  [arXiv:1509.00834 [hep-ph]].
  
\bibitem{Matsedonskyi:2015xta} 
  O.~Matsedonskyi,
  JHEP {\bf 1601}, 063 (2016)
  doi:10.1007/JHEP01(2016)063
  [arXiv:1509.03583 [hep-ph]].
  
\bibitem{Evans:2016htp} 
  J.~L.~Evans, T.~Gherghetta, N.~Nagata and Z.~Thomas,
  JHEP {\bf 1609}, 150 (2016)
  doi:10.1007/JHEP09(2016)150
  [arXiv:1602.04812 [hep-ph]].
  
\bibitem{Lalak:2016mbv} 
  Z.~Lalak and A.~Markiewicz,
  arXiv:1612.09128 [hep-ph].
  
\bibitem{Batell:2017kho} 
  B.~Batell, M.~A.~Fedderke and L.~T.~Wang,
  arXiv:1705.09666 [hep-ph].
  
\bibitem{Nelson:2017cfv} 
  A.~Nelson and C.~Prescod-Weinstein,
  arXiv:1708.00010 [hep-ph].
  
\bibitem{Huang:2016dhp} 
  F.~P.~Huang, Y.~Cai, H.~Li and X.~Zhang,
  Chin.\ Phys.\ C {\bf 40}, no. 11, 113103 (2016)
  doi:10.1088/1674-1137/40/11/113103
  [arXiv:1605.03120 [hep-ph]].
  
\bibitem{Fowlie:2016jlx} 
  A.~Fowlie, C.~Balazs, G.~White, L.~Marzola and M.~Raidal,
  JHEP {\bf 1608}, 100 (2016)
  doi:10.1007/JHEP08(2016)100
  [arXiv:1602.03889 [hep-ph]].
  
  
\bibitem{Hook:2016mqo} 
  A.~Hook and G.~Marques-Tavares,
  JHEP {\bf 1612}, 101 (2016)
  doi:10.1007/JHEP12(2016)101
  [arXiv:1607.01786 [hep-ph]].
  
\bibitem{You:2017kah} 
  T.~You,
  arXiv:1701.09167 [hep-ph].
  
  \bibitem{Tangarife:2017rgl} 
  W.~Tangarife, K.~Tobioka, L.~Ubaldi and T.~Volansky,
  arXiv:1706.03072 [hep-ph].
  
\bibitem{DiChiara:2015euo} 
  S.~Di Chiara, K.~Kannike, L.~Marzola, A.~Racioppi, M.~Raidal and C.~Spethmann,
  Phys.\ Rev.\ D {\bf 93}, no. 10, 103527 (2016)
  doi:10.1103/PhysRevD.93.103527
  [arXiv:1511.02858 [hep-ph]].
  
\bibitem{Marzola:2015dia} 
  L.~Marzola and M.~Raidal,
  Mod.\ Phys.\ Lett.\ A {\bf 31}, 1650215 (2016)
  doi:10.1142/S0217732316502151
  [arXiv:1510.00710 [hep-ph]].
  
\bibitem{Choi:2015fiu} 
  K.~Choi and S.~H.~Im,
  JHEP {\bf 1601}, 149 (2016)
  doi:10.1007/JHEP01(2016)149
  [arXiv:1511.00132 [hep-ph]].
  
\bibitem{Kaplan:2015fuy} 
  D.~E.~Kaplan and R.~Rattazzi,
  Phys.\ Rev.\ D {\bf 93}, no. 8, 085007 (2016)
  doi:10.1103/PhysRevD.93.085007
  [arXiv:1511.01827 [hep-ph]].
  
\bibitem{McAllister:2016vzi} 
  L.~McAllister, P.~Schwaller, G.~Servant, J.~Stout and A.~Westphal,
  arXiv:1610.05320 [hep-th].
  
\bibitem{Fonseca:2016eoo} 
  N.~Fonseca, L.~de Lima, C.~S.~Machado and R.~D.~Matheus,
  Phys.\ Rev.\ D {\bf 94}, no. 1, 015010 (2016)
  doi:10.1103/PhysRevD.94.015010
  [arXiv:1601.07183 [hep-ph]].

\bibitem{Dvali:2003br} 
  G.~Dvali and A.~Vilenkin,
  Phys.\ Rev.\ D {\bf 70}, 063501 (2004)
  doi:10.1103/PhysRevD.70.063501
  [hep-th/0304043].

\bibitem{Dvali:2004tma} 
  G.~Dvali,
  Phys.\ Rev.\ D {\bf 74}, 025018 (2006)
  doi:10.1103/PhysRevD.74.025018
  [hep-th/0410286].

\bibitem{Broy:2015qna} 
  B.~J.~Broy, M.~Galante, D.~Roest and A.~Westphal,
  JHEP {\bf 1512}, 149 (2015)
  doi:10.1007/JHEP12(2015)149
  [arXiv:1507.02277 [hep-th]].

\bibitem{Scalisi:2016hvu} 
  M.~Scalisi,
  arXiv:1607.01030 [hep-th].

\bibitem{Choi:2016kke} 
  K.~Choi, H.~Kim and T.~Sekiguchi,
  Phys.\ Rev.\ D {\bf 95}, no. 7, 075008 (2017)
  doi:10.1103/PhysRevD.95.075008
  [arXiv:1611.08569 [hep-ph]].


  
\bibitem{Anber:2009ua} 
  M.~M.~Anber and L.~Sorbo,
  Phys.\ Rev.\ D {\bf 81}, 043534 (2010)
  doi:10.1103/PhysRevD.81.043534
  [arXiv:0908.4089 [hep-th]].
  

  
  \bibitem{Ferreira:2017lnd} 
  R.~Z.~Ferreira and A.~Notari,
  arXiv:1706.00373 [astro-ph.CO].
  
\bibitem{Bellac:2011kqa} 
  M.~L.~Bellac,
  ``Thermal Field Theory''
  
\bibitem{Kapusta:2006pm} 
  J.~I.~Kapusta and C.~Gale,
  ``Finite-temperature field theory: Principles and applications''
  
\bibitem{Linde:1980ts} 
  A.~D.~Linde,
  Phys.\ Lett.\  {\bf 96B}, 289 (1980).
  doi:10.1016/0370-2693(80)90769-8
  


  
\end{thebibliography}
\end{document}